\documentclass[conference]{IEEEtran}
\IEEEoverridecommandlockouts

\usepackage{cite}
\usepackage{amsmath,amssymb,amsfonts}
\usepackage{algorithmic}
\usepackage{textcomp}

\usepackage[english]{babel}
\usepackage[utf8]{inputenc}
\usepackage{microtype}
\usepackage{url}
\usepackage{csquotes}
\usepackage{listings}
\usepackage{nth}

\usepackage[disable]{todonotes}

\usepackage{booktabs}
\usepackage[bookmarks=false]{hyperref}
\usepackage[capitalize]{cleveref}

\usepackage{graphicx}
\usepackage{xcolor}
\graphicspath{ {./graphics/} }
\DeclareGraphicsExtensions{.pdf}

\usepackage{enumitem}
\newlist{paraenum}{enumerate*}{1}
\setlist[paraenum]{label=(\emph{\roman*})}

\usepackage[nohyperlinks]{acronym}

\acrodef{ADC}{Analogue-Digital Converter}
\acrodef{CAN}{Controller Area Network}
\acrodef{ECU}{Electronic Control Unit}
\acrodef{MAC}{Message Authentication Code}
\acrodef{RPM}{Revolutions Per Minute}
\acrodef{USB}{Universal Serial Bus}

\begin{document}

\title{Automated Fuzzing of Automotive Control Units}

\author{
\IEEEauthorblockN{Timothy Werquin,
        Roos Hubrechtsen,
        Ashok Thangarajan,
        Frank Piessens and
        Jan Tobias M\"uhlberg}
\IEEEauthorblockA{\textit{KU Leuven, Dept. of Computer Science,
imec-DistriNet} \\ B-3001 Leuven, Belgium \\
  \url{<firstname.lastname>@cs.kuleuven.be}}
}

\maketitle

\begin{abstract}
Modern vehicles are governed by a network of Electronic Control
Units~(ECUs), which are programmed to sense inputs from the driver and the
environment, to process these inputs, and to control actuators that, e.g.,
regulate the engine or even control the steering system. ECUs within a
vehicle communicate via automotive bus systems such as the Controller Area
Network~(CAN), and beyond the vehicles boundaries through upcoming
vehicle-to-vehicle and vehicle-to-infrastructure channels.  Approaches to
manipulate the communication between ECUs for the purpose of security
testing and reverse-engineering of vehicular functions have been presented
in the past, all of which struggle with automating the detection of system
change in response to message injection. In this paper we present our
findings with fuzzing CAN networks, in particular while observing
individual ECUs with a sensor harness. The harness detects physical
responses, which we then use in a oracle functions to inform the fuzzing
process. We systematically define fuzzers, fuzzing configurations and
oracle functions for testing ECUs. We evaluate our approach based on case
studies of commercial instrument clusters and with an experimental
framework for CAN authentication. Our results show that the approach is
capable of identifying interesting ECU states with a high level of
automation. Our approach is applicable in distributed cyber-physical
systems beyond automotive computing.

\end{abstract}

\begin{IEEEkeywords}
automotive control networks, CAN, security, testing, fuzzing
\end{IEEEkeywords}

\section{Introduction}
Modern cars are largely controlled by software. This software forms a
distributed mixed-criticality system that executes on a number of
interconnected \acp{ECU}. Jointly, these \acp{ECU} govern the vehicle's
behaviour -- from convenience and infotainment functions to safety-critical
functionality.  \acp{ECU} are connected via automotive bus systems that
facilitate the exchange of messages, most of which communicate sensor
readings and control instructions. The control software then interprets
these messages and reacts to events by triggering the relevant actuators,
e.g., brakes, airbags, or steering gear.  In 2016, the Ford Motor Company
reported that their latest models are running on 150 million lines of code.
Given the enormous complexity of these systems, they are notoriously hard
to test, for safety as well as for security properties.

Since around 2004, researchers have been expressing their concerns with respect to
the security limitations of communication standards, including the widely
used \ac{CAN}~\cite{wolf2004auto_bus_sec,hoppe2008can_security,henninger2009sec_requirements}. Since 2010, a series of
high-profile
attacks~\cite{koscher2010experimental,checkoway2011comprehensive,miller2014survey,miller2015remote_car_exploit} illustrate that with increased
vehicular connectivity even remote adversaries can take control of
critical functions of a vehicle. These risks have been acknowledged and are
partly addressed in
emerging industry standards~\cite{sae:j3061,autosar42} that encompass
authentication and software security for control systems, and prototypes
that showcase secure system designs for automotive computing based on
software attestation and Trusted Computing
primitives~\cite{vanbulck:sancus-automotive} have been proposed.
Meanwhile, more and more low-level vulnerabilities in these communication
systems are being revealed (e.g., \cite{froeschle2017can-attacker} and
\cite{palanca:2017-stealth-can}), guidance for the reverse-engineering
and penetration testing of vehicular communications and control systems
becomes readily available~\cite{smith:2016-car-hackers-handbook}, and the
need for advanced testing methodology for these systems is generally
acknowledged. A testing approach that promises a particularly high level
of automation is fuzzing.

Fuzz testing~\cite{oehlert:2005-fuzzing,manes:2019-fuzzing} is a
well established methodology to expose software and systems to unexpected
conditions, for example by providing random input streams that may crash
the target. Yet, the approach does not easily apply to embedded
software~\cite{muench:2018-fuzzing} and few approaches have been made to
fuzz embedded control systems or automotive \acp{ECU} in
particular~\cite{checkoway2011comprehensive,lee:2015-can-fuzzing,bayer:2016-automotive}. A key difficulty to overcome here is the
definition of oracle functions that define when a fuzzer has potentially
triggered a bug or at least an \enquote{interesting} system state, and to automatically
evaluate these functions. 

\subsubsection{Our Contributions.}
In this paper we discuss fuzz testing in the context of automotive control
networks. Specifically, our research investigates the use and automation of
fuzzing so as to find vulnerabilities and to reverse engineer \ac{ECU}
functionality in \ac{CAN} networks. We make the following contributions:

\begin{enumerate}
  \item{We systematically define fuzzers, fuzzing configurations and
    oracle functions for testing automotive \acp{ECU} through their
    \ac{CAN} interface.}
  \item{We develop a sensor harness to automatically evaluate fuzzing
    oracles for \acp{ECU} with physical outputs.}
  \item{We evaluate our approach, taking commercial automotive instrument
    clusters and an experimental setup for testing AUTOSAR-compliant
    message authentication as case studies.}
\end{enumerate}

To the best of our knowledge, this paper is the first to
largely automate a methodical fuzzing approach (e.g.
following~\cite{manes:2019-fuzzing}) for automotive \acp{ECU}.  Although
our implementation is targeting \ac{CAN} components, our approach can be
generalised to cyber-physical systems with any underlying communication
technology. Our fuzzer implementation, instructions to build the sensor
harness and to repeat our experiments are available under an
open-source license at
\href{https://github.com/timower/caringcaribou/tree/autoFuzz}{https://github.com/timower/caringcaribou/tree/autoFuzz}.

\section{Background}
In this section we briefly introduce the \ac{CAN} bus, which is commonly
used to facilitate communication between automotive \acp{ECU} but also in
industrial control systems. We further introduce the CaringCaribou
penetration testing framework, which our fuzzing toolchain is integrated
with.

% -------------------------------------------------------------------------
\subsection{Controller Area Network (CAN) \& Security}
\begin{figure}[h]
    \centering
    \includegraphics[width=.95\linewidth]{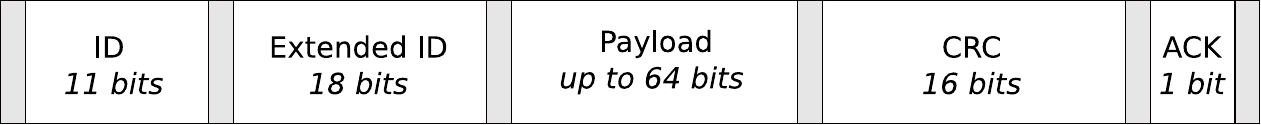}
    \caption{Extended data frame standardised by CAN 2.0B.}
    \label{fig:can_frame}
\end{figure}

The \ac{CAN} bus is the most commonly used broadcast network in modern
cars.  A \ac{CAN} message consists of an 11-bit arbitration ID, followed by
an optional 18-bit extended ID, and up to 8 bytes of data payload (cf.
Fig.~\ref{fig:can_frame}).  Dedicated transceiver hardware implements a
protocol for message acknowledgement and bus arbitration for
sending/receiving data frames.  \ac{CAN} requires a fixed data transmission
rate, and allows recessive bits (one) to be overwritten by dominant bits
(zero) during transmission.  Message acknowledgement can thus simply be
implemented by overwriting the ACK bit at the end of the data frame in
real-time.  Likewise, to implement bus arbitration, \ac{CAN} transceivers
are required to continuously listen on the bus while sending the message ID
at the beginning of the data frame, and to back off when one of their ID
bits has been overwritten.  This scheme ensures that messages with lower
IDs effectively have higher priorities. Finally, each \ac{CAN} frame
features a 16-bit CRC field to detect transmission errors.

\ac{CAN} was originally developed in 1983, when cyber attacks were of no
concern. Thus, the protocol does not provide any form of message
authentication.  Any \ac{ECU} connected to the network can spoof messages
with arbitrary sender ID and payload, which forms the basis of many
attacks~\cite{koscher2010experimental,checkoway2011comprehensive,miller2014survey,miller2015remote_car_exploit}.
As a response, the AUTOSAR~\cite{autosar42} standardisation body published
industry guidelines for backwards-compatible message authentication in
vehicular networks. The VulCAN framework~\cite{vanbulck:sancus-automotive},
which we study in Sect.~\ref{sect:vulcan} is one implementation of these
authentication extensions.

% -------------------------------------------------------------------------
\subsection{CaringCaribou and Automotive Penetration Testing}
We built our implementations of \ac{CAN} fuzzers as modules for the
open-source penetration-testing framework CaringCaribou. Our fuzzing
extensions are freely available for further experimentation and follow-up
research.  Since fuzzers are widely used as a means to perform black-box
testing in the regular penetration testing industry, it is our belief that
a similar tool could prove useful in the automotive penetration testing
community.

CaringCaribou\footnote{\url{https://github.com/CaringCaribou/caringcaribou}}
is a tool developed for the purpose of being the \enquote{nmap} of
automotive security.  Cyber security research in the automotive industry is
a new field that is rapidly expanding.  Yet, it still lacks the mature
tooling available to the mainstream security community.  Tools like
CaringCaribou aim to fill that gap.  CaringCaribou has a modular
architecture that allows developers of penetration testing techniques for
automotive systems to easily write new modules for their specific purpose,
and deploy these modules using a unified tooling infrastructure.  Thus,
CaringCaribou provides the developer with a layer of abstraction which
protects them from the specifics of \ac{CAN} and other automotive
communication protocols, allowing them to focus on writing the actual
penetration testing tool instead of dealing with the lower-layer
interactions.  CaringCaribou is supposed to be a zero-knowledge tool that
can be deployed on any \ac{CAN} network regardless of its specific
configuration.

\section{Fuzzing CAN Networks}
In this section we lay out our approach to define a fuzzing tool for
\ac{CAN}-based automotive control systems. We follow the approach of Manes
et al.~\cite{manes:2019-fuzzing} and dissect the tool into an oracle
component, the actual fuzzer and the run-time configuration for the fuzzer
for a particular run.

% -------------------------------------------------------------------------
\subsection{Bug Oracles for ECUs} \label{sect:oracles}
According to~\cite{manes:2019-fuzzing}, a (bug) oracle is a program that
determines whether a given execution of the target system violates a
specific (security) policy. In Sect.~\ref{sect:eval} we outline two very
different case studies for our system: In one of these we have a partial
specification of security properties of the network available, and where we
are looking for violations of this specification. In the second case study
we have no reliable specification but we are interested in
reverse-engineering such a specification. Both case studies are
characterised by not being able to observe software interactions directly
(as opposed to software fuzzing with code instrumentation in a
simulator~\cite{manes:2019-fuzzing}). Instead, we are looking at black-box
systems to which our fuzzer can provide an input stream, while any
observable communication output, physical output (actuation of a LED or a
relay) or even the timing or absence of such outputs (e.g., due to a
software crash) may indicate that an \enquote{interesting} system state has
been reached.

Recent related work in the field of intrusion detection for industrial
control system (e.g.,~\cite{taylor:2016-anomaly} and~\cite{wresnegger:2018-anomaly}) suggests that machine-learning approaches
can be used to train detectors that then report anomalies in the
communication behaviour of vehicular networks. We have not implemented such
oracle functions.

Physical outputs of control units can certainly be observed by human
operators. They can also be sensed and electronically reported through
sensor networks, or in our case a sensor harness that is attached to the
target system. In the following sections we emphasise on this form fuzzing
oracle, where a state change in the target is defined by a sensor
(de-)activation or sensor threshold.

Most difficult is certainly the detection of system failure which results
in the absence of an observable response from the target. Thus, inputs that
lead to failures are easily misinterpreted by a fuzzing tool as inputs that
have no effect. For example, our work deals with \acp{ECU} that need to
regularly receive certain messages or they will fault, effectively rendering
the continuation of a fuzzing campaign ineffective. Detecting these
messages and constructing traffic that satisfies the input requirements of
\acp{ECU} in the absence of a specification, is difficult. An oracle to
identify these kinds of system faults requires the use of recorded traffic
that periodically triggers a known observable output. An oracle function
will then fire when the periodic signal is absent due to, e.g., message
omission. We apply this method in our \emph{omission fuzzer} below.

% -------------------------------------------------------------------------
\subsection{Defining CAN Fuzzers}
Fuzzing is the execution of the target system using input(s) sampled from
an input space (the \enquote{fuzz input space}) that protrudes the expected
input space of the target system~\cite{manes:2019-fuzzing}. With fuzzing we
aim to enumerate and exercise a large subset of this fuzz input space to
find system behaviour that triggers an oracle function. \acp{ECU} that
process \ac{CAN} messages are an interesting target since the frame size of
\ac{CAN} messages is at most 110~bits. This fuzz input space is certainly
huge, but much smaller than, e.g., Ethernet frames, WiFi frames, or
multimedia streams.  Still, even for \ac{CAN} networks, this fuzz input
space is prohibitively large for being exhaustively exercised. Furthermore,
with a maximum bandwidth of 1~MBit, and most \acp{ECU} using 500~MBit as a
fixed transfer rate, data transmission to a target network of \acp{ECU}
represents a bottleneck.

Starting with the idea of \emph{random fuzzing}, where arbitration IDs and
message payloads are selected randomly, we devise three additional fuzzing
strategies, \emph{brute-force fuzzing}, \emph{mutation fuzzing} and
\emph{identify fuzzing}, to narrow down the fuzz input space and explore
interesting \ac{ECU} behaviour more efficiently. These strategies are based
on the observation that an \ac{ECU} typically accepts inputs on a
relatively small number of IDs only, that also the number of payload bits
that result in a observable state change is limited, and that several
consecutive messages may be required to trigger an observable state change.
We then integrate these approaches in an \emph{automated exploration} mode,
where inputs from a sensor harness (cf. Sect.~\ref{sect:sensors}), which is
attached to a target \ac{ECU}, guide input generation. We have implemented
our approach in two modules for CaringCaribou, namely \texttt{fuzzer} and
\texttt{autoFuzz}, which can be invoked as \texttt{./cc.py <module>
<parameters> [-f <file>]}. Here, \texttt{./cc.py} refers to the
CaringCaribou main script, \texttt{<module>} to a fuzzer module, and
\texttt{<parameters>} to a fuzzer configuration which we discuss below.
\texttt{-f <file>} can be used to store a message trail on disk. For
example, \texttt{./cc.py fuzzer random} will generate entirely random
messages and dispatch them over the configured \ac{CAN} interface.

\subsubsection{Brute-Force Fuzzing.} This method aims to exhaustively
enumerate selected hexadecimal digits in a message, specifically in the
message's ID field and the payload. For example, the fuzzer can be invoked
as \texttt{./cc.py fuzzer brute 0x123 12ab..78}, where the \nth{5} and
\nth{6} octet of the message payload will be enumerated and sent, while the
message ID \texttt{0x123} and all other payload octets remain constant.

\subsubsection{Mutation Fuzzing.} This strategy can be used to
systematically explore a larger fuzz input space through mutating selected
hex digits in arbitration ID and message by means of individual random bit
flips. An example use for this strategy is \texttt{./cc.py fuzzer mutate
7f.. 12ab....}; the syntax follows the example given for \emph{brute-force
fuzzing} above.

\subsubsection{Identify Fuzzing.} Once a fuzzing run resulted in an event
of interest, e.g., a change of an indicator LED on a target \ac{ECU}, the
\emph{identify} method can be used to replay and identify a minimal set of
messages that caused the event. The syntax for invoking this method is
\texttt{./cc.py fuzzer identify log.txt}, where \texttt{log.txt} refers to
a log file previously recorded with the \texttt{-f} parameter. The method
relies on human input -- i.e., key presses -- to gather information about
the timely occurrence of events, and aims to prune the set of recorded
messages in \texttt{log.txt} so that the event still occurs when the pruned
set is replayed.

\subsubsection{Automated ECU Exploration.} Our \texttt{autoFuzz} module
implements the above strategies so that system change can be detected
directly through our sensor harness (cf. Sect.~\ref{sect:sensors}). Sensor
observations can then be used to guide the generation of the next inputs
and to automatically identify message bits that lead to observable system
change, depending on the fuzzing strategy. The module further features the
generation of J1939-compliant messages and the fuzzing of J1939 function
group addresses (PGNs and SPNs).

When fuzzing an \ac{ECU} with a single sensor attached to one of the
\ac{ECU}'s actuators, it is possible to immediately run the identify fuzzer
when a change in the sensor state is detected without relying on recorded
traffic. This requires that the actuator can be triggered with a
predictable payload.  When using multiple sensors,  the log file can be
filtered to keep a number of messages preceding the activation of a
specific sensor which can then be used as input to the identify fuzzer.

While experimenting with fuzzing strategies, we observed that an \ac{ECU}'s
response to a message is often delayed.  During the delay period, other
messages are being sent by the fuzzer, which makes identifying the CAN
messages specifically responsible for an response more difficult.  One
possible solution is to increase the delay between sending messages. Yet,
this will increase the time required to cover the fuzz input space.
Another option is the resend only a subset of the messages preceding a
sensor activation with increased delays, which we implemented in our
identify method.

Our experiments further revealed that some \acp{ECU} expect certain
messages to be received regularly. The absence of these messages will lead
to a shut-down or render the \ac{ECU} unresponsive and to indicate a
failure. These behaviours prevent our identify method from working as
sending the complete traffic log can keep the \ac{ECU} responsive but
sending parts of the recorded traffic will cause the \ac{ECU} to fault. To
address this, we developed an approach that we refer to as \emph{omission
fuzzing}. This strategy sends the complete recorded traffic but omits some
messages in order to identify which message cause specific state changes.
The identified arbitration IDs or payloads can then be added to a
\enquote{blacklist} to inform other methods. E.g., any arbitration ID in
the blacklist will never be omitted during \emph{identify fuzzing}.

After collecting logs from the fuzzer, various analysis can be applied. For
example, if a specification of the target \ac{ECU} is available, which would
detail the expected actuator responses to specific groups of messages,
sensor activations can be check against this specification to detect bugs
or undocumented behaviour.

% -------------------------------------------------------------------------
\subsection{Target-Specific Fuzzer Configuration}

A fuzz configuration of a fuzzer comprises the parameter value(s) that
control(s) the fuzz algorithm~\cite{manes:2019-fuzzing}. In the context of
our approach, these parameters involve \emph{message generation},
\emph{message timing}, \emph{message omission}, and the configuration of
the \emph{sensor harness}. As outlined before fuzzing entire \ac{CAN}
messages makes little sense as it results in an extremely large fuzz input
space. Thus, configurations will typically restrict the fuzz space to
specific octets in (extended) arbitration IDs and message payload. Message
timing is typically configured to schedule a new message every 3~ms to
20~ms to avoid message collisions and to leave enough time for actuators to
be engaged and sensed. Message omission and baseline traffic are to be set
up to simulate typical bus traffic in a car so as to make target \acp{ECU}
function normally. The sensor harness offers a wide range of configuration
options that involve the type of sensors, sampling rates, the number of
sensors and their placement on the target \ac{ECU}.

\section{A Sensor Harness to Automate ECU Fuzzing} \label{sect:sensors}
In this section we describe an inexpensive and extensible experimental
sensor harness to automate the analysis of automotive \acp{ECU}. The
intuition behind the setup is that fuzzing communication in an automotive
control network, or in cyber-physical systems in general, can cause a range
of interesting responses beyond network communication. Thus, to use these
responses as inputs to fuzzing oracles (cf. Sect.~\ref{sect:oracles}) they
must be automatically measured at an appropriate sampling rate. Previous
approaches to consider these responses typically rely on human observation
and human interaction during the fuzzing process. For example, a fuzzing
tool may require the user to press a key if they observe a change in the
system, e.g., a flashing indicator light on a control panel.  Our work
improves over this by detecting physical responses of \acp{ECU}
automatically, with negligible delays, and at a configurable granularity.

\begin{figure}
  \centering
  \includegraphics[width=.95\linewidth]{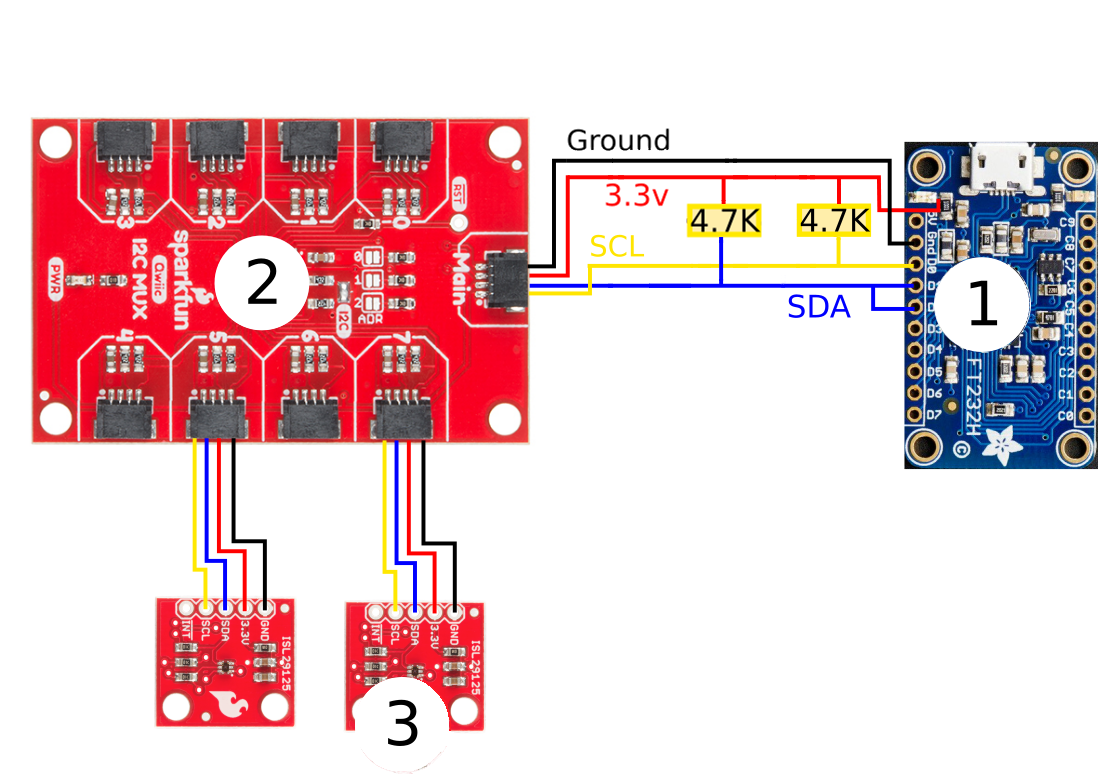}
  \caption{Sensor harness connection schema showing \emph{(1)} the FT232H
  I2C-to-USB converter, \emph{(2)} an TCA9548A I2C Multiplexer, and
  \emph{(3)} two ISL29125 light \& colour sensors.}
  \label{fig:sensors-schema}
\end{figure}

Fig.~\ref{fig:sensors-schema} gives an overview of the sensor harness. The
system in action is depicted in Sect.~\ref{sect:eval-cluster}. The harness
connects multiple sensors together and provides a \ac{USB} interface for a
PC to control the setup. The depicted configuration contains only light and
colour sensors which can be placed over the various indicators LEDs on a
target \ac{ECU}.  No general-purpose microcontroller is used in the
harness, which allows the entire setup to be programmed and configured from
the PC in a higher-level language, Python in our case.

The light sensors are connected through an I2C bus, a communication
interface which is present on many low-cost sensors. The use of I2C make
the harness extensible with various other sensors, such as sound sensors to
monitor auditory alerts, motion sensors to monitor steering wheel movement,
or current sensors to detect an engine start.
As the I2C bus supports multiplexing, multiple sensors can be connected to
the same bus as long as they have a different I2C address.
The low-cost sensors used in the harness often have a fixed address, which
implies that an I2C multiplexer must be used to connect multiple sensors of the
same type in the harness. In order to interface with the sensor harness
from a PC an USB to I2C adapter is used.

% -------------------------------------------------------------------------
\subsubsection{Light \& Colour Sensors: ISL29125.}
The ISL29125 colour sensor is used to measure the status of visual
indicators on automotive equipment.  The sensor provides a simple RGB light
level readout and has a number of configuration options which determine the
sensitivity and precision of the sensor.  The sensitivity can be configured
between 10k~lux or 375~lux and the sensor has a built in infrared light
filter which can be configured separately.  The precision of the sensor can
be configured to be either 12 or 16 bits.  Changing the precision also
changes the integration time, meaning a higher precision (16 bits) will
require the \ac{ADC} to sample the sensor longer resulting in slower
measurements.  In addition to the I2C interface the sensor has an interrupt
pin which can be triggered by a configurable light level on either of the
red, green or blue channel. Currently the harness does not use the
interrupt functionality but works by polling each sensor individually.  Our
colour sensors have a single fixed I2C address, requiring a multiplexer to
connect multiple sensors on the same I2C bus.

% -------------------------------------------------------------------------
\subsubsection{I2C Multiplexer: TCA9548A.}
As the colour sensor has a fixed I2C address, the TCA9548A multiplexer
is used to connect multiple colour sensors in the harness.  The TCA9548A
multiplexer has eight I2C channels which allows eight colour sensors to be
connected on the same bus.  As the multiplexer has a three bit configurable
address multiple multiplexers can be daisy chained allowing up to 64 I2C
busses on a single connection.

% -------------------------------------------------------------------------
\subsubsection{I2C-to-USB: FT232H.}
In order to connect the I2C bus to a PC the FTDI FT232H adapter is used.
This adapter supports a number of different bus protocols such as UART, SPI
and I2C in addition to a GPIO interface which can be used to write to eight
digital IO pins.

% -------------------------------------------------------------------------
\subsubsection{Programming, Calibration \& Use.}
To obtain readings from the light sensors, the Arduino
library\footnote{\url{https://github.com/sparkfun/SparkFun_ISL29125_Breakout_Arduino_Library}}
for reading ISL29125 sensors was adapted.  The resulting library uses the
Adafruit\footnote{\url{https://github.com/adafruit/Adafruit-FT232H-Breakout-PCB}}
library for communicating with the FT232H adapter.  Our current sensor
library exposes functions for initialising a new sensor, and for reading
the sensor's red, green and blue values.

The initialisation function resets the sensor and configures it in RGB mode
which enables all three colour channels, puts it in 10k-lux mode for bright
environments and enables high IR light adjustment.  Reading a colour value
is done by reading from the relevant device register, which returns a
colour value of either 12 or 16 bits depending on the chosen precision.
The time the sensor needs to take a reading varies depending on the
\ac{ADC} integration time which in turn depends on the chosen precision: At
16 bit precision, each reading takes about 110~ms while at 12 bit each
reading takes only 7~ms.  As the sensor's output registers are
double-buffered, reading out these registers between sensing operations
will result in outdated readings.

In order to use multiple light sensors a library that interfaces with the
TCA9548A multiplexer was created, this library allows virtual I2C ports to
be created for each channel of the multiplexer.  These virtual I2C ports
can then be used in the sensor library instead of the default FT232H I2C
port.  In order to switch I2C channels the number of the requested channel
is written before any commands, this adds a delay before every I2C command,
which is negligible in comparison with the integration time of the light
sensor and does not impact fuzzing performance.

In initial experiments we use the colour light sensors to detect whether the
various indicators on an automotive dashboard are changing state, effectively
converting the red, green and blue light levels into a binary input signal for
the fuzzer.  As the sensors are sensitive enough to detect (even reflected)
movements behind the sensor while duct-taped to a dashboard in both sensitivity
configurations (up to 375~lux and 10k~lux) a simple threshold is not sufficient
to distinguish state change. We devise a calibration method that involves taking
a reading when the indicator is on and when the indicator is off.  This results
in red, green and blue light level triples to which any new measurement can be
compared, if the new measurement is closer to the on-value the indicator is
detected as on and vice versa.  This method assures that a uniform increase or
decrease in ambient light does not change the detected indicator value.  The
method further requires a calibration with the indicator both on and off, which
may not be feasible when the indicator trigger is unknown.  When the indicator
cannot be triggered during calibration, a simple threshold may be used to detect
the indicator state.  More elaborate calibration methods may be required to
operate the sensor harness in noisy environments.

\section{Evaluation and Discussion} \label{sect:eval}
We have applied our fuzzer implementation and the sensor harness to a
number of case studies that include the ICSim automotive instrument cluster
simulator\footnote{\url{https://github.com/zombieCraig/ICSim}}, a demo
setup for illustrating and implementing message authentication in \ac{CAN}
networks with the VulCAN~\cite{vanbulck:sancus-automotive} framework, as
well as real instrument clusters. In this section we focus on our
experience and lessons learned from the latter two case studies. We compare
our findings with earlier manual approaches to discover bugs and explore
proprietary functionality in these scenarios.

% -------------------------------------------------------------------------
\subsection{Case Study 1: VulCAN} \label{sect:vulcan}
We evaluated the effectiveness of our fuzzer to find implementation bugs
and security vulnerabilities on a demo implementation of
VulCAN~\cite{vanbulck:sancus-automotive}, a generic design for \ac{CAN}
message authentication. VulCAN provides efficient and
AUTOSAR-compliant~\cite{autosar42} authentication plus software component
attestation based on lightweight trusted computing technology.  We used the
same test bench as described in~\cite{vanbulck:sancus-automotive} to test
the abilities of the fuzzer.

In brief, the demo consists of a number of \acp{ECU} with keypads as input
devices and LED displays as actuators. A distributed control application
which simulates a traction control system is executing on the \acp{ECU}.
Application components communicate via cryptographically authenticated
\ac{CAN} messages with freshness guarantees: only messages that are
successfully validated to be fresh and to originate from unmodified and
integrity-protected remote component should ever be able to trigger output
events. The application communicates only a few valid payloads at fixed
intervals. Thus, deviation from expected behaviour would be easy to detect.
Yet, since it is unlikely for a random or mutation-based fuzzer to
\enquote{guess} a valid payload, nonce, and authentication tag triple, and
since the system was designed with security in mind, we did not expect the
security properties of the system to be broken easily.  This part of the
evaluation is conducted without using the sensor harness but by relying on
visual observation on the demo's LED displays.  The fuzzer is executing on
desktop PC, which is connected to the demo setup via a \ac{USB} to \ac{CAN}
interface.

To our surprise, with the help of the fuzzer, we detected and traced
several unique vulnerabilities in the system in a fairly short period of
time. Below we focus on two particularly subtle discoveries.

The first vulnerability was discovered nearly instantaneously in a fuzzer
configuration where messages with extended \ac{CAN} arbitration IDs are
generated. Such messages resulted in system states where the injected
messages could lead to actual display outputs, breaking the security
properties of VulCAN entirely. Extended \ac{CAN} IDs are not being used in
the VulCAN demo, and thus, the components were not tested in environments
where these messages occur. Most likely, a misconfigured driver for the \ac{CAN}
controller on an \acp{ECU} -- \enquote{untrusted} software in VulCAN's
attacker model -- together with an incomplete rejection condition in a
secure application module, allowed an attacker to arbitrarily adjust the
displays of the test bench without having to pass authenticity checks.

The second vulnerability was found within the implementation of one of the
authentication protocols in VulCAN, specifically
VatiCAN~\cite{nurnberger2016vatican}. This implementation turned out to be
particularly vulnerable to denial-of-service attacks when being flooded
with specific traffic patterns, allowing an attacker to desynchronise
nonces and render trusted components unresponsive even to dedicated
re-synchronisation messages. The bug was discovered in a timespan of
several minutes when fuzzing the test bench in a configuration where both,
the fuzzer as well as an \ac{ECU}, are simultaneously attempting to send
messages to a target \ac{ECU}. Interestingly, due to the configuration
error in \ac{CAN} drivers described above, messages with extended \ac{CAN}
IDs are effectively interpreted as broadcast messages. Application
components are thus subject to receiving a mix of fuzzer-generated payloads
and authenticated messages which results in denial-of-service.

% -------------------------------------------------------------------------
\subsection{Case Study 2: Instrument Clusters}
\label{sect:eval-cluster}
In the context of automotive security research and for building demos such
as the VulCAN~\cite{vanbulck:sancus-automotive} setup, instrument clusters
are commonly used as easily accessible off-the-shelf components with many
visible indicators (speed needle, turning indicators, display, etc.), most
of which can be controlled through \ac{CAN} messages. Yet, the specific
arbitration IDs and payloads to control these functions are not
publicly documented. Literature on car hacking (e.g.,
\cite{smith:2016-car-hackers-handbook}) suggests manual approaches to
reverse-engineer these details, which may require hours or even days of
try-and-error, even for a skilled engineer. By using our sensor harness, we
expect a substantial speed-up of these processes, on top of being able to
largely automate the process.
 
\begin{figure*}
  \centering
  \includegraphics[width=.58\linewidth]{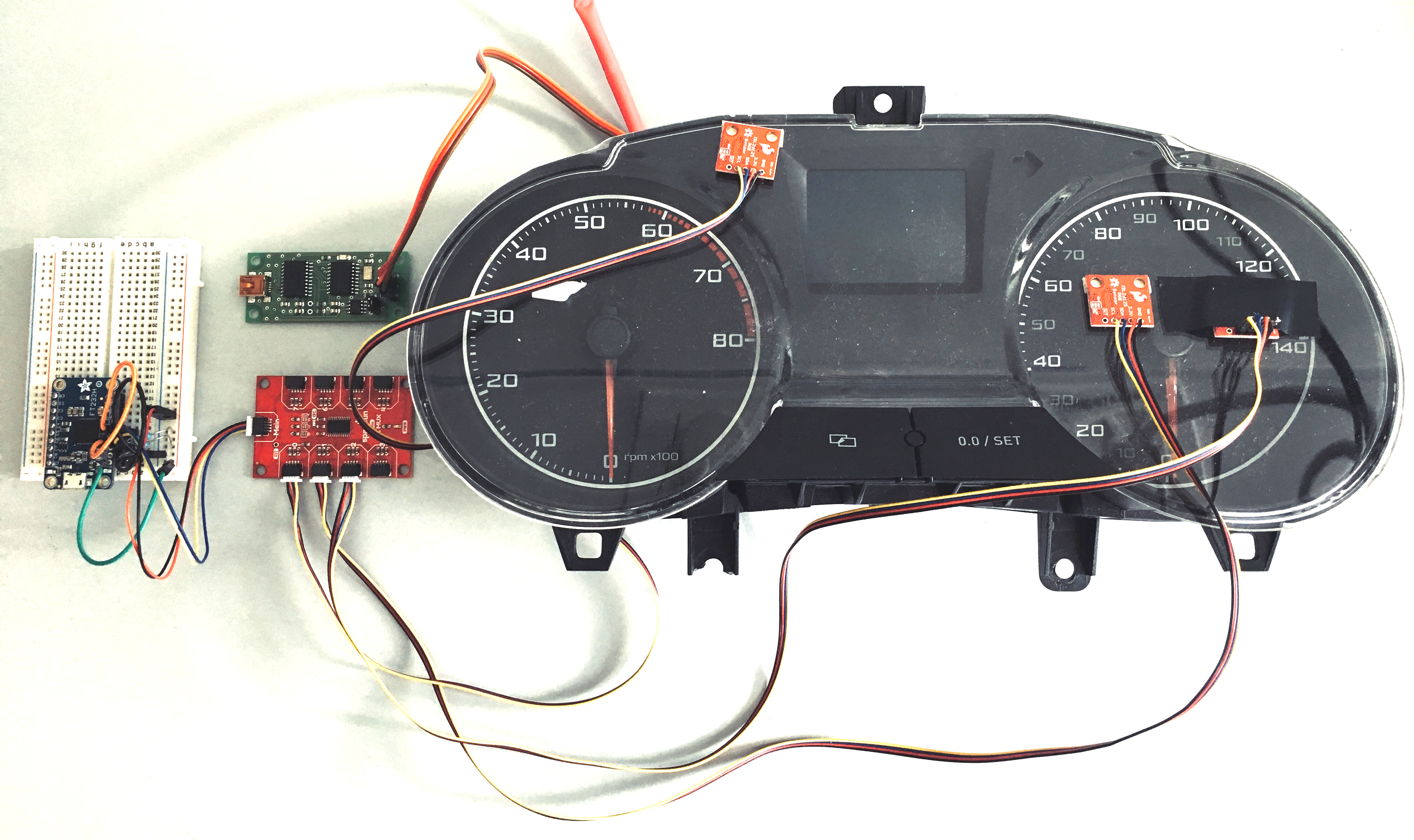}
  \caption{An automotive instrument cluster with (part of) our sensor harness
    attached. The cluster originates from a 2014 Seat Ibiza model.}
  \label{fig:sensors-setup}
\end{figure*}
 
We have been experimenting with a number of clusters from passenger cars
and commercial vehicles. As illustrated in Fig.~\ref{fig:sensors-setup},
components of our sensor harness are duct-taped to indicators of the
cluster. The instrument cluster is connected to a desktop PC with the
fuzzer via a \ac{USB} to \ac{CAN} interface and there are no other
\acp{ECU} present on the \ac{CAN}.

In order to test the instrument cluster's basic functions, we developed a
controller application that would send a number of
documented~\cite{dashboardracesim} \ac{CAN} messages to the dashboard. Only
some these control messages worked in combination with our dashboard. We
then applied our fuzzer in \emph{identify} mode to filter the traffic data
for messages that trigger physical functions, then applied
\emph{brute-force} and \emph{mutation} mode to explore arbitration IDs and
payloads to trigger further functionality. 

By \emph{brute-forcing} the entire 11-bit arbitration ID fuzz-space with a
fixed payload \texttt{0xffffffff}, most indicators LEDs in the dashboard
could be activated.  Some of these indicators are switched on by default
when the instrument cluster is powered up. These indicators could be
triggered using a fixed payload of \texttt{0x00000000}.  Control of the
speedometer and engine RPM needles could also be triggered. The fuzzer
takes about 30~s to enumerate the 11-bit address space using a delay of
10~ms between messages.  Running the \emph{identify} method after finding a
state changes takes an extra 30~s per activation in order to identify the
message responsible for an indicator activation.

Using \emph{mutation-based} fuzzing on any of the messages identified
with the brute-force method, we were able to reverse engineer the semantics
of most payload bits.  For example, by starting from the messages that
activated the left turning indicator, the fuzzer was able to not only
identify the bit responsible for triggering the indicator. It was also able
to identify the bit responsible for the right turn signal, the headlights
indicator and a number of other status LEDs.  Enumerating eight payload
bits takes about 20~s with a delay of 3~ms between messages. Using the
\emph{identify} method requires an additional 5~s per indicator.  The delay
between messages is critical when using mutation-based fuzzing.  If the
delay is too short, the next message, which will have another bit flipped,
will overwrite the previous message.  This will cause more indicator
activations to be missed or some indicators to not even activate.

\emph{Omission fuzzing} is not necessary for analysing this particular
instrument cluster. Yet, we heavily relied on this method method when
working with more modern clusters from commercial vehicles.

% -------------------------------------------------------------------------
\subsection{Discussion \& Lessons Learned}
The two case studies outlined above show that our fuzzers can efficiently
reveal undocumented functionality, intricate bugs and security
vulnerabilities in \acp{ECU} connected to \ac{CAN} networks. Ultimately,
our experiments provide further evidence for fuzzing to be a useful tool in
testing and reverse-engineering, which is due to the technique's ability to
cover an enormous range of possible combinations of system inputs, in our
case arbitration IDs and payloads. Many of these inputs may not even occur
in normal and benign operation, and are difficult to consider in static
test cases.

Above we describe two critical security vulnerabilities in an experimental
system design, which are based on intricate implementation and
configuration bugs. Previously undetected, these vulnerabilities became
apparent within minutes with the use of a fuzzer, even without relying on a
sensor harness. The harness could be used to reduce human interaction and
to improve the duration for detecting and tracing these bugs, but
non-trivial extensions of the harness would be required to sense the state
of actuators such as the attached displays. Alternatively, the demo setup
could be modified to feature simpler actuators (i.e., individual LEDs or
relays) that allow for an easier detection of conditions that satisfy our
bug oracles. We used our fuzzer, specifically the identify and replay
functionality, to trace bugs in source code and fix vulnerabilities. The
resulting fixes are not straight-forward as they require consideration of
rather involved network states. Our findings highlight the need for
thorough testing and verification on top of strong cryptographic primitives
and Trusted Computing technology when designing distributed control systems
that are potentially exposed to malicious interactions.

We further described how the sensor harness in combination with our fuzzing
techniques can be used to largely automate the process of
reverse-engineering communication protocols of proprietary \acp{ECU}.
Manually reversing a substantial subset of the functionality of, e.g., an
instrument cluster, can easily be an effort of several days or even weeks.
With our approach, this can be achieved within hours. Additional sensors
(e.g., audio, power consumption, vibration) could further extend the
harness' abilities. We believe that our approach can be used to identify
bugs and unintended functionality when being applied to components for
which a specification is available. This specification could be integrated
in a bug oracle such that responses outside of the specified behaviour are
detected as errors. In this context, our approach may be useful to automate
activities such as integration testing and to achieve a high input-space
coverage in these activities.

Fuzzing \acp{ECU} under realistic conditions, i.e., while being connected
to a vehicle's \ac{CAN} network with many other \enquote{noisy} \acp{ECU},
may also be feasible but requires fuzzing strategies that aim at noise
reduction by exploring the effects of individual messages or sequences of
messages in different system states. In this context it may be useful to
also consider \ac{CAN} responses of \acp{ECU}. We may borrow from recent
approaches in anomaly detection in control systems (e.g.,
\cite{taylor:2016-anomaly} and \cite{wresnegger:2018-anomaly}) to define
oracle functions that detect changes in the response stream of an
individual \ac{ECU}, or even to detect state change throughout the network.

\section{Related Work}
Fuzzing has a long history and is still actively developed, in particular
in the domain of security- and penetration testing of software systems~\cite{oehlert:2005-fuzzing,manes:2019-fuzzing}. Recent
work~\cite{muench:2018-fuzzing} elaborates on the difficulties of employing
fuzzing in embedded systems. Specifically for automotive systems, Smith
states in~\cite{smith:2016-car-hackers-handbook} that, while fuzzing can
certainly be useful in discovering undocumented services or crashes, it is
rarely useful beyond that, e.g., to find and exploit vulnerabilities. Our
experience report disagrees slightly with this observation: We discovered
that fuzzing is more efficient in finding subtle vulnerabilities and
configuration errors than monitoring or reverse engineering the firmware
and communications. Fuzzing exposes substantially more of the system's
unintended states than what one would be able to explore manually, due to
the sheer amount of pseudo-random message combinations that are generated
and dispatched by the fuzzer.  This allows testers to focus on tracking
down and responding to vulnerability reports instead of having to manually
probe the system.  With automated oracle function, as discussed in
Sect.~\ref{sect:oracles} and Sect.~\ref{sect:sensors}, fuzzing becomes even
more efficient. While our approach mostly relies on black-box fuzzing where
very little knowledge of the system is assumed and oracle functions must
rely on system outputs rather than observing the system's internal state,
our approach can certainly be combined with more advanced
reverse-engineering and firmware inspection tools. This would lead to more
powerful and also much more intricate oracle functions.

Related research investigates the extent to which fuzzing can be applied be
in automotive systems~\cite{lee:2015-can-fuzzing,bayer:2015-car-fuzzing,nishimura:2016-canfd-fuzzing,fowler:automating-car-fuzzing,bayer:2016-automotive}.  Our
work aims to improve over this state of the art by not only defining a
fuzzer for \ac{CAN} networks, but by developing an entire methodology that
defines fuzzing objectives, oracle functions and fuzzing strategies, and
substantially improves the automation of testing cyber-physical systems.
We report on experiments and lay out our experience from applying this
methodology to two realistic systems, one of which being a prototype for an
automotive security system. While related work reports mixed results on the
usefulness of fuzzing automotive networks, we judge our results as largely
positive since we were able to identify a few subtle vulnerabilities and
dramatically speed up reverse-engineering activities.

Further related research investigates the use of fuzzing to explore
stateful communication protocols~\cite{banks:2006-snooze,gorbunov:2010-autofuzz}, also in the context of cyber-physical
systems such as smart-grid communications~\cite{dantas:2014-efuzz}. Our
toolchain does currently not employ such techniques. We believe that
these could be a sensible addition to the current stateless exploration
approach. Specifically, these techniques may be used to brute-force
intricate \ac{CAN} protocols that trigger, e.g., software updates on
\acp{ECU}.

\section{Summary \& Conclusions}
Automotive control networks are highly complex safety-critical and
security-critical systems which have been shown to be vulnerable to
adversarial interactions. In this paper we devise a largely automated
approach for fuzz-testing these systems. We discuss how bug oracles for
automotive \acp{ECU} can be described, define a number of fuzzing
strategies, and develop a sensor harness to allow oracle functions to
detect interesting system behaviour in an automated fashion. We have
implemented our fuzzing approach in CaringCaribou, an open-source
automotive penetration-testing toolkit, and we report on two sets of
experiments where we apply our fuzzer to find vulnerabilities and to
reverse-engineer proprietary \ac{ECU} functions. To the best of our
knowledge, our approach is the first to achieve a high degree of automation
for these activities, and we see future applications of our fuzzing
approach in, e.g., penetration testing but also in integration- and
compliance testing. While we have been focusing on \ac{CAN} networks, we
believe that the approach is applicable to other types of control networks
and beyond the domain of automotive computing. In the future we will work
towards a more rigorous evaluation of our approach, following the
methodology of~\cite{klees:2018-evaluating-fuzzing}.

\subsubsection{Acknowledgements.}
This research is partially funded by the Research Fund KU Leuven. This
research is partially funded under SErVO, \enquote{Secure and Economically
Viable V2X Solutions}, by the Flemish Agentschap Innoveren \& Ondernemen.
We thank the developers of CaringCaribou for their sport and ideas, and
for integrating parts of our fuzzer into their platform.

\bibliographystyle{IEEEtran}
\bibliography{bibliography}

\end{document}